\documentclass[10pt,final,journal,twocolumn]{IEEEtran}
\usepackage{amsmath}
\usepackage{graphicx, epstopdf}
\usepackage{setspace}
\usepackage{acronym}
\usepackage{commath}
\usepackage{mathrsfs}
\usepackage{mathtools}
\usepackage{calc}
\usepackage[top=1in, left=0.7in, right=0.7in]{geometry}
\usepackage{ifthen}
\usepackage{textcomp}
\usepackage{float}
\usepackage[tight,footnotesize]{subfigure}
\usepackage{cite}
\usepackage{bm}
\usepackage{url}
\usepackage[utf8]{inputenc}
\usepackage{mathtools}
\usepackage{amssymb}
\usepackage{amsthm}
\usepackage{mathrsfs}
\usepackage{bm}
\usepackage{multirow}
\usepackage{multicol}
\usepackage{xspace}
\usepackage{tabularx}

\usepackage{cite}
\usepackage{amsmath,amssymb,amsfonts}
\usepackage{algorithmic}
\usepackage{algorithm}
\usepackage{graphicx}
\usepackage{textcomp}

\usepackage{xcolor}
\def\BibTeX{{\rm B\kern-.05em{\sc i\kern-.025em b}\kern-.08em
    T\kern-.1667em\lower.7ex\hbox{E}\kern-.125emX}}

\begin{document}

\title{Agentic Assistant for 6G: Turn-based Conversations for AI-RAN Hierarchical Co-Management}

\author{Udhaya Srinivasan$^{1}$, Weisi Guo$^{1*}$
\thanks{$^{1}$All authors are with Cranfield University, Bedford, UK. $^{*}$Corresponding: weisi.guo@cranfield.ac.uk. 
This work is supported by EPSRC CHEDDAR: Communications Hub For Empowering Distributed ClouD Computing Applications And Research (EP/X040518/1, EP/Y037421/1).}
}

\maketitle

\begin{abstract}
New generations of radio access networks (RAN), especially with native AI services are increasingly difficult for human engineers to manage in real-time. Enterprise networks are often managed locally, where expertise is scarce. Existing research has focused on creating Retrieval-Augmented Generation (RAG) LLMs that can help to plan and configure RAN and core aspects only. Co-management of RAN and edge AI is the gap, which creates hierarchical and dynamic problems that require turn-based human interactions. Here, we create an agentic network manager and turn-based conversation assistant that can understand human intent-based queries that match hierarchical problems in AI-RAN. The framework constructed consists of: (a) a user interface and evaluation dashboard, (b) an intelligence layer that interfaces with the AI-RAN, and (c) a knowledge layer for providing the basis for evaluations and recommendations. These form 3 layers of capability with the following validation performances (average response time 13s): (1) design and planning a service (78\% accuracy), (2) operating specific AI-RAN tools (89\% accuracy), and (3) tuning AI-RAN performance (67\%). These initial results indicate the universal challenges of hallucination but also fast response performance success that can really reduce OPEX costs for small scale enterprise users.
\end{abstract}


\section{Introduction}
Enterprise radio-access networks (RAN) have quietly become the invisible fabric of the digital economy. The AI-RAN fabric orchestrates a wide range of sensors and actuators, from tele-operating cranes, flying drones, to factory robots \cite{10148954}. Yet the management interface is decades old and telecom engineers still analyse log fragments and handcraft YAML snippets (e.g., "Why is sector 3 dropping packets?" or "How can we balance workload between these edge GPUs?"). Most enterprises highlighted integration complexity, troubleshooting delays, and the need for enhanced analytics as principal operational challenges in enterprise RAN \cite{Omdia}. The burden on human cognitive load is not merely an inconvenience for small companies; it is a hidden cost multiplier that delays Industry 4.0 roll-outs and forces some enterprises to over-provision spectrum and compute.

\subsection{Review of AI in Planning and Management}

Our early work has focused on building agentic AI to help design and plan networks, especially when there is a gap between customer requirements and understanding of native-AI and RAN capabilities \cite{RAG6G}. This produced intent-based calls to a RAG-LLM pipeline that extracted knowledge vectors from 3GPP standards and highly rated academic papers and industrial reports, improving the likelihood of valid designs. A feedback loop was added to allow customers to vote on specific knowledge vectors from documents. It can provide designs for both AI-for-RAN \cite{Nvidia} (e.g., RRM \cite{RRM} or sensing \cite{RFSensingGPT}), and RAN-for-AI (e.g., RAN provides for distributed/federated AI \cite{FLGPT25}). 

In terms of real-time management, emerging research has focused on using knowledge-graph based RAG to help create contextualized and trustworthy processes for network management \cite{Graph}, with interactive agents that maintain a human-in-the-loop insight \cite{AgentInterface}. Furthermore, current evaluation approaches focus largely on one-off requests \cite{AgentBench, 10758700} and do not address hierarchical turn-based conversations.

\subsection{Gap and Novelty}
Very few current work has shown how we can jointly co-manage RAN and native-AI resources together, and this is important for digital transformation because edge intelligence is what orchestrates resources. This is difficult, because it is a hierarchical problem and require the joint consideration of both AI compute for RAN and RAN to facilitate AI compute distribution \cite{Nvidia}. As such, rather than a one-off question from the human user \cite{AgentBench, 10758700}, our work needs to tackle turn-based conversations that has to be assessed over the whole process.

Here, our innovation proposes:
\begin{itemize}
    \item Designing turn-based conversational agents that can speak to both telecom experts and end-users for both RAN-for-AI and AI-for-RAN service queries \cite{Nvidia}. These are designed to decompose hierarchical problems, follow temporal dynamics, and have multi-modal feedback.
    \item Designing a knowledge layer that can engage both the RAN emulator and conversation agents to provide knowledge
    \item Validate the conversational agents to systematic scrutiny via HATT-E (Hierarchical Agent Task \& Tool Evaluation), a bespoke agentic evaluation protocol designed for AI-RAN problems comprising 50 ecologically valid, multi-turn interaction traces representative of routine operational tasks. 
\end{itemize}
The analysis provides the early work for other researchers to develop operational capable AI assistants to manage not only the RAN, but also native AI within the RAN.

\section{System Setup and Methodology}

\begin{figure}[!t]
\centering
\includegraphics[width=3in]{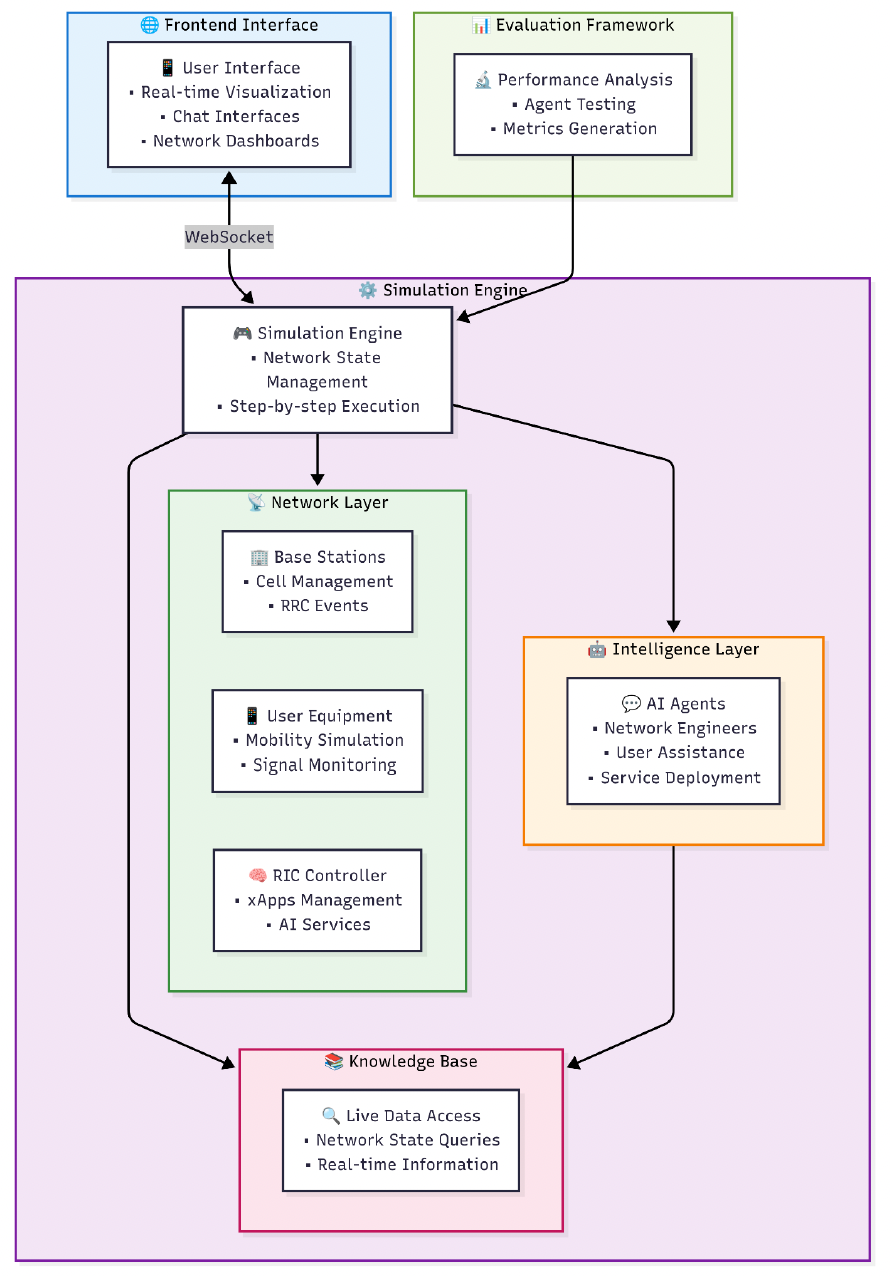}
\caption{System diagram of agentic network management and network simulator.}
\label{figs1}
\end{figure}

An overall system diagram is shown in Fig.~\ref{figs1} and we breakdown the 3 layers below. Our setup is open sourced and can be found in footnotes \footnote{AI-RAN-Sim: an AI-enabled O-RAN simulation platform: https://github.com/ntutangyun/ai-ran-sim.}.

\subsection{AI-RAN Emulator for ORAN}
\begin{figure*}[!t]
\centering
\includegraphics[width=6in]{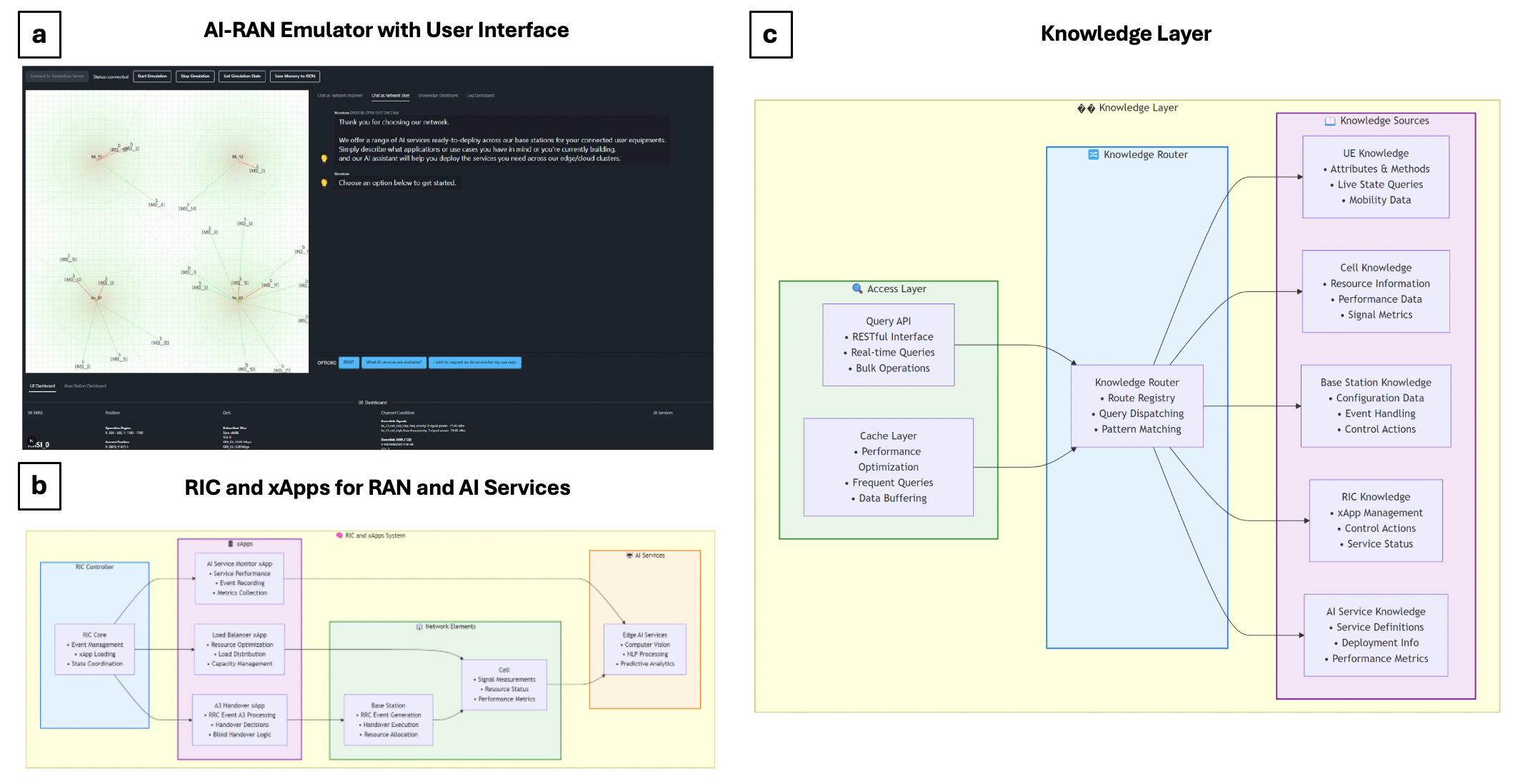}
\caption{System modules: (a) AI-RAN emulator that mirror BubbleRAN ORAN, (b) RIC and xApp ecosystem for RAN and AI service control, and (c) Knowledge layer for agentic AI conversation agents.}
\label{figs2}
\end{figure*}

Whilst there is an abundance of RAN emulators that support packet-level behaviour, load patterns, and control-plane signaling (e.g., OpenNet, Simu5G), they lack human-in-the-loop (HITL) interfaces and native-AI capability. A good example is EdgeCloudSim, but still use a YAML-based operator centric interface system and is not well suited to enterprise customer engagement. Across them, the common limitation is one of a script-bound interaction with no conversation agents. As such, they are not well suited to emerging use cases and enterprise customers. This is why we have used our own emulator built in-house to do AI-RAN emulations that mirrors our BubbleRAN ORAN \cite{10758700}, which comprises of (see Fig.~\ref{figs2}a):
\begin{itemize}
    \item base stations: event processing, handover, and radio resource management (RRM)
    \item environment: signal propagation, interference modeling
    \item edge AI server: AI model hosting, container management, compute provisioning
    \item user equipment (UE): mobility, application traffic
    \item Radio Intelligent Controller (RIC) \& control plane: xApps and management, load balancing, event subscription, lifecycle management
    \item Core network: slice management, authentication
\end{itemize}
The \textit{RIC} component represents the intelligent control layer of the O-RAN architecture (see Fig.~\ref{figs2}b), managing xApps and AI service subscriptions. The RIC dynamically loads and manages xApps, which are modular applications that implement specific network control and optimization functions. The RIC implements an event-driven architecture where xApps can subscribe to network events and issue control actions in response. The RIC also manages AI service subscriptions, enabling dynamic deployment and scaling of AI services based on network demand and user requests.

\subsection{Intelligence Layer}
The intelligence layer provides AI-powered capabilities for network management, user assistance, and service deployment. This includes both edge computing infrastructure to deploy AI services for user applications (RAN-for-AI) and improving RAN performance (AI-for-RAN) \cite{Nvidia}. The AI service management system supports various types of AI services, including computer vision, natural language processing, and predictive analytics. Each service is containerized and can be deployed on edge servers associated with base stations, enabling distributed AI processing throughout the network.

There is provision for an AI service pipeline implements a two-stage process for AI service deployment:
\begin{enumerate}
    \item Profiling: where the system analyses user requirements and recommends appropriate AI services. 
    \item AI Deployment: including resource allocation, container deployment, and service configuration. 
\end{enumerate}
The pipeline includes sophisticated algorithms for resource optimization (details in our earlier work \cite{AutoML25}), ensuring that AI services are deployed efficiently while maintaining network performance and meeting user requirements.

\subsection{Knowledge Layer}
The \textit{Knowledge Router} serves as the central access point for all knowledge queries, implementing a routing system that directs queries to appropriate knowledge sources (see Fig.~\ref{figs2}c). The router supports pattern-based matching, enabling flexible querying of network data and simulation state. The knowledge router maintains a registry of available knowledge sources and their capabilities, enabling efficient query processing and resource utilization. The router also implements caching mechanisms to improve query performance for frequently accessed data. 

Knowledge sources are specialized components that provide access to specific types of network information. Each knowledge source implements standardized interfaces for data access and query processing, ensuring consistency and interoperability across the system. The knowledge sources cover all major network elements, including user equipment, base stations, cells, RIC components, and AI services. Each source provides both current state information and historical data, enabling comprehensive analysis and decision-making.

\begin{figure*}[!t]
\centering
\includegraphics[width=6in]{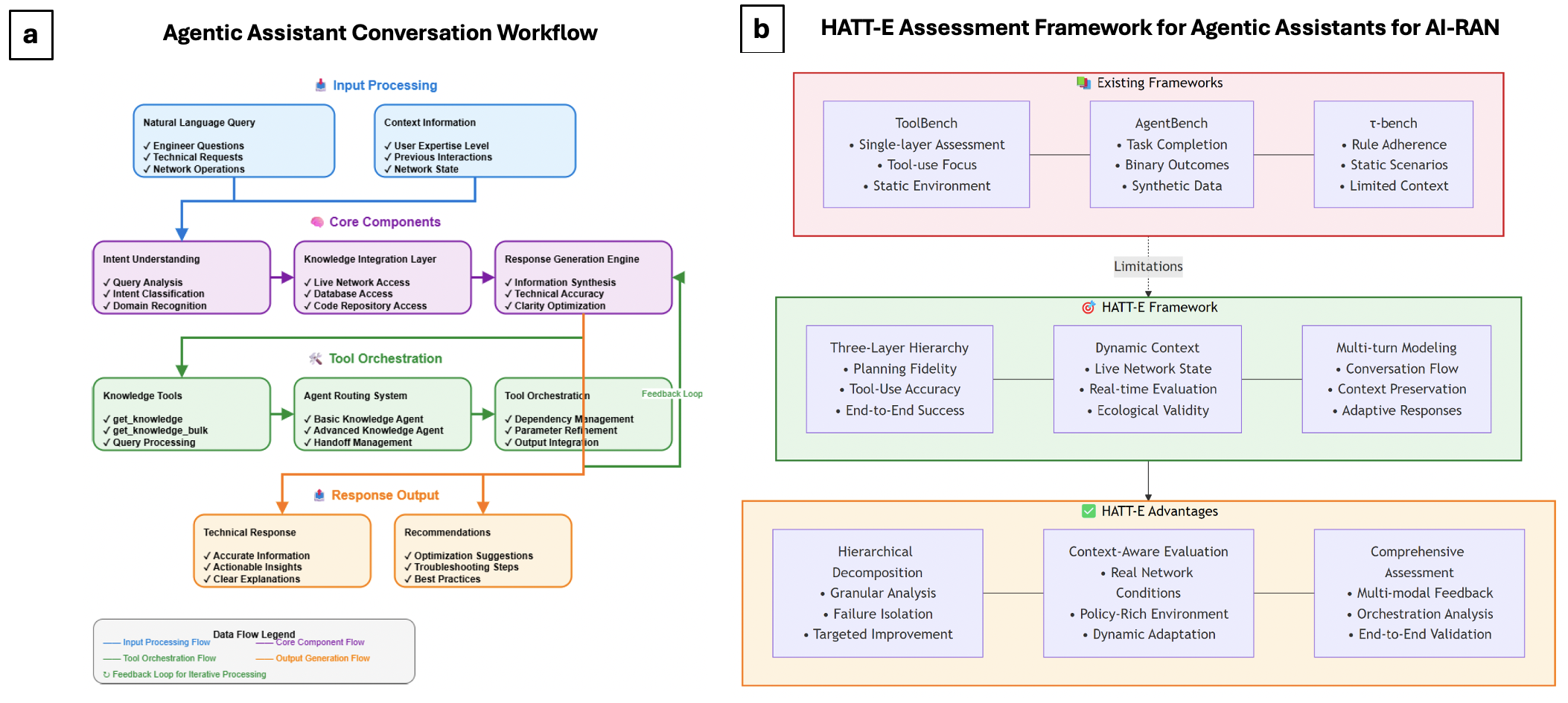}
\caption{Design of Conversational Agentic Assistant and Assessment Framework.}
\label{figs3}
\end{figure*}

\subsection{Interface Layer}
The front-end interface is built using Next.js, a modern React-based framework that provides excellent performance and developer experience. This choice enables the creation of responsive, interactive web applications that can handle real-time data updates and complex user interactions efficiently. The front-end layer is designed to serve multiple user roles, including network engineers, system administrators, and researchers.

\subsubsection{Conversation Agentic Assistant: EngineerChat} EngineerChat (see Fig.\ref{figs3}a) serves as an AI-powered assistant for network engineers, providing expert guidance on network configuration, troubleshooting, and optimisation. Building upon the intent-driven networking concepts demonstrated in recent O-RAN research and extending the collaborative automation approach proposed by Maestro \cite{10758700}, the agent is architected as a hierarchical system with three primary components: (1) Intent Understanding, (2) Knowledge Integration, and (3)
Response Generation.

\subsubsection{Conversation Agentic Assistant: UserChat} UserChat targets non-technical subscribers and system administrators, enabling natural-language requests for AI service deployment and network resource management. The agent abstracts complex network operations behind conversational interfaces, democratizing access to edge AI capabilities. This approach extends the end-to-end edge AI service provisioning concepts demonstrated in recent 6G O-RAN research by us \cite{RAG6G}, providing a human-friendly interface layer. This agent analyses user requirements expressed in natural language and maps them to appropriate AI service configurations. This engine also uses the knowledge layer to access the database of available AI models, their resource requirements, and deployment constraints. The models are maintained as docker image in a private docker repository, from where the models can be easily pulled and deployed on the edge-AI nodes.

\section{HATT-E Assessment Framework for Conversational Agentic Assistants}
The integration of conversational AI agents into complex radio-access network environments presents unique evaluation challenges that existing benchmarks fail to address. While conventional AI evaluation frameworks such as ToolBench and AgentBench \cite{AgentBench} focus on isolated task completion or tool-use accuracy, they do not capture the hierarchical nature of network operations nor the multi-layered decision-making processes inherent in radio-edge environments. This section presents the design and implementation of HATT-E (Hierarchical Agent Task \& Tool Evaluation), a novel evaluation framework specifically to assess conversational agents (see Fig.~\ref{figs3}c). 

\subsection{AI-RAN Unique Design Requirements}
The unique characteristics of radio-access network operations impose specific requirements on evaluation frameworks:
\begin{itemize}
    \item Hierarchical Decision Decomposition: operations typically involve multiple levels of abstraction, from high-level intent through task decomposition to specific tool usage. 
    \item Temporal Dynamics: states evolve continuously, making static ground-truth datasets inadequate. 
    \item Multi-Modal Feedback: synthesize information from multiple sources (telemetry, configuration databases, log streams) and present coherent analyses to operators. Evaluation must assess both the accuracy of individual data retrievals and the quality of synthesized insights.
\end{itemize}

\begin{figure*}[!t]
\centering
\includegraphics[width=6in]{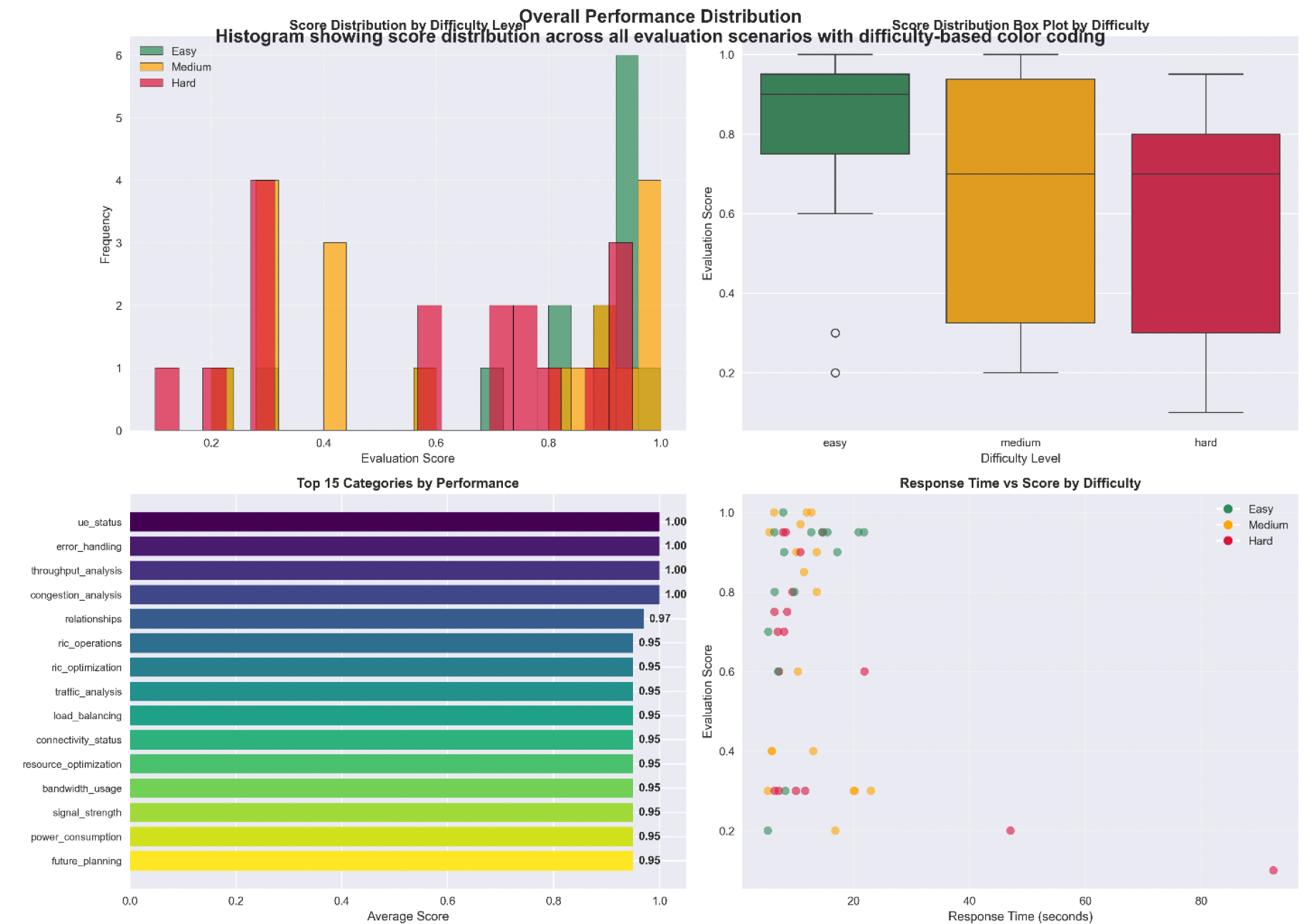}
\caption{Overall Performance Distribution Results across 50 Evaluation Scenarios with Difficulty-based Colour Coding.}
\label{figs4}
\end{figure*}

\subsection{AI-RAN informed HATT-E Framework}
HATT-E implements a three-layer evaluation hierarchy that isolates different aspects of agent performance.

Layer 1: Planning Fidelity evaluates whether agents correctly decompose high-level user queries into appropriate sub-tasks. This layer measures the agent’s understanding of network operations and its ability to structure complex queries into manageable components. Planning fidelity is assessed by comparing the agent’s task breakdown against expert-defined reference plans for each scenario.

Layer 2: Tool-Use Accuracy measures the precision with which agents select and invoke individual network analysis tools. This layer captures whether agents choose appropriate tools for specific sub-tasks. Tool-use accuracy is quantified through parameter validation and output verification against expected results.

Layer 3: End-to-End Task Success assesses whether the complete user query is satisfied, regardless of the internal planning or tool-use strategies employed. This layer provides the ultimate measure of utility from the operator’s perspective and captures emergent behaviors that arise from the interaction between planning and execution layers.

\subsection{Ground Truth for Validation}
Establishing reliable ground truth for conversational network agents presents unique challenges, as network operations often admit multiple valid solution paths. HATT-E addresses this through a hybrid approach combining automated verification with AI-based assessment:
\begin{itemize}
    \item Automated Verification for Deterministic Answers: directly compares against simulation ground truth.
    \item AI-Based Assessment for Complex Queries: evaluated against consensus between multiple AI agents using semantic similarity metrics.
    \item Multi-Path Validation: The framework recognizes that network troubleshooting may follow different valid approaches. Agent responses are evaluated for technical validity rather than exact path matching, allowing for innovation in problem-solving approaches while maintaining accuracy standards.
\end{itemize}
The multi-turn conversation evaluation employed a comprehensive dataset of \textbf{50 scenario-based question sets} spanning \textbf{3 difficulty levels} (easy, medium, hard) with balanced coverage across 50 distinct operational categories. Each evaluation scenario involved dynamically generated conversations of 2-3 turns, reflecting realistic network troubleshooting and service deployment workflows. 

\section{Evaluation Results}

\subsection{Overall Results}
The evaluation (see Fig.~\ref{figs4}) revealed solid overall performance with an average score of 0.6724 across all scenarios. Performance demonstrated graceful degradation with increasing difficulty: easy scenarios achieved 0.7933, medium scenarios scored 0.6428, and hard scenarios attained 0.5971. This pattern indicates stable agent behavior under varying complexity levels while highlighting opportunities for improvement in complex operational tasks. \textit{Tool usage analysis} revealed a strong preference for precise, scoped knowledge retrievals over bulk operations. The agents made 136 calls compared to only 8 calls, reflecting efficient single-entity queries but potential missed opportunities for comparative analysis tasks. 

The \textbf{Key Performance Indicators} are summarized below: 
\begin{itemize}
    \item Overall average score: 0.6724 
    \item Response time: 13.35 seconds average
    \item Tool efficiency: 94\% single-entity queries vs 5.6\% bulk operations 
    \item Difficulty scaling: 25\% performance degradation from easy to hard scenarios.
\end{itemize}

\begin{figure}[!t]
\centering
\includegraphics[width=3in]{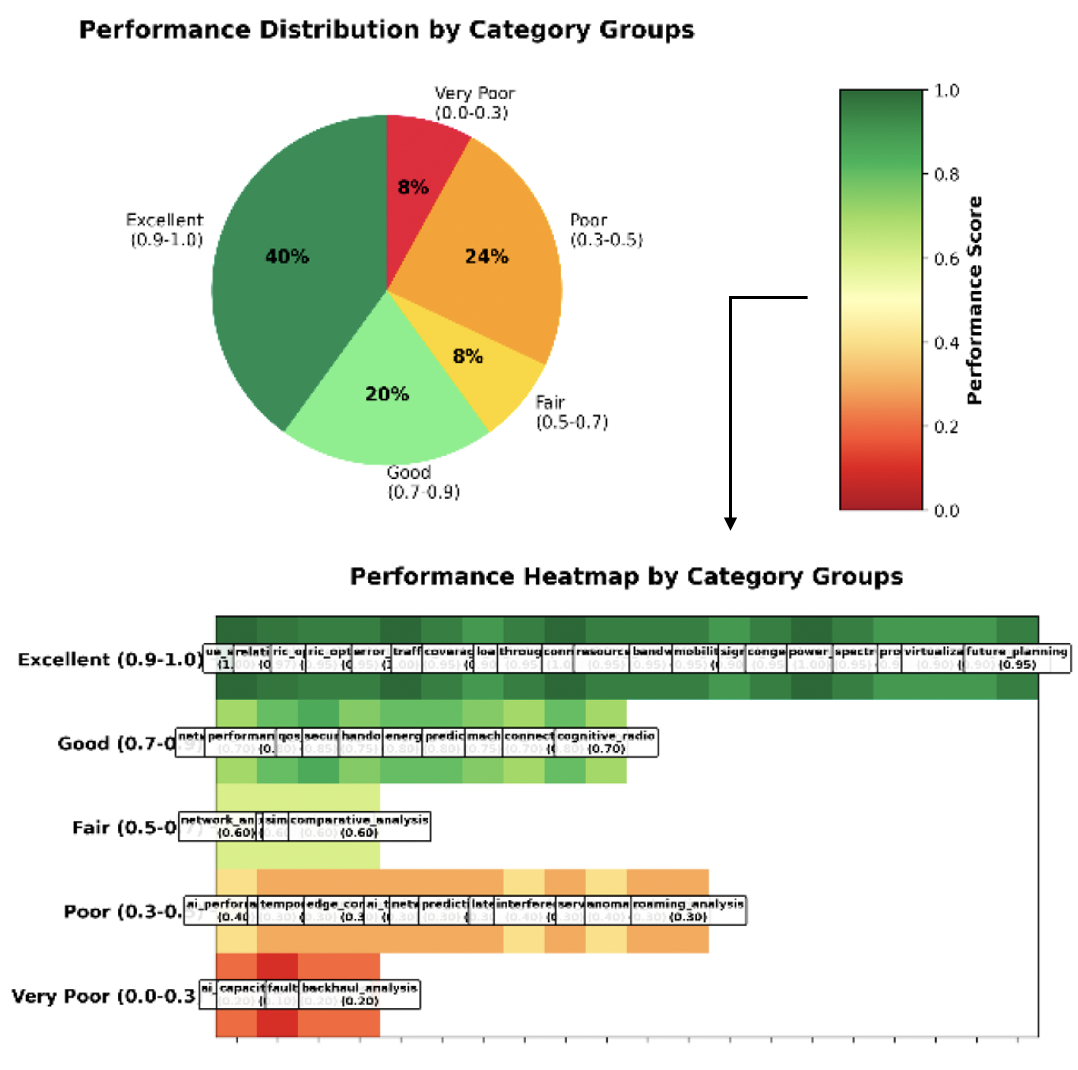}
\caption{Category Results over 50 Operational Scenarios.}
\label{figs5}
\end{figure}

\subsection{Category Results over 50 Scenarios}
Performance analysis across operational categories revealed distinct patterns of strength and weakness (see Fig.\ref{figs5}). Top-performing categories included fundamental network operations such as UE status monitoring (1.00), throughput analysis (1.00), and congestion analysis (1.00). These categories benefit from direct mapping to single knowledge endpoints and straightforward data interpretation. Conversely, low-performing categories clustered around complex synthesis tasks: capacity planning (0.10), backhaul analysis (0.20), AI services management (0.20), and fault prediction (0.20). These categories require multi-signal integration and predictive analysis capabilities that exceed current agent capabilities.

\subsection{Summary Discussion}
Our turn-based agentic conversation analysis (HATT-E) laid bare the specific weaknesses that contribute to this performance gap. At the \textit{planning layer}, a poor delegation accuracy of 48\% revealed a fundamental inability to correctly decompose problems and route them to the appropriate specialized function. This is a critical failure point in any multi-agent or tool-augmented system, as it indicates the agent lacks a robust internal model of the problem domain and its own capabilities. The consequences of this were evident in the 41 documented cases of redundant planning steps, which directly impact efficiency and cost. Even more concerning was the finding at the \textit{execution layer}. A high tool success rate of 85\% suggesting technical competence was undermined by a hallucination rate of 43\%. This means that in nearly half of the interactions, the agent generated factually incorrect or unsubstantiated information, even when the underlying tools were used correctly. 

\section{Conclusions and Future Work}
Enterprise AI-RAN need agentic assistants especially when small businesses struggle to pay for experts. Here, we designed turn-based conversation agents that can hierarchically plan, deploy, and manage AI-for-RAN and RAN-for-AI services \cite{Nvidia}. The principal conclusion is while agents demonstrate near-perfect proficiency in routine fact-retrieval tasks, they struggle with complex reasoning, synthesis, and prediction tasks. The overall performance score of 0.67, while respectable, masks an underlying variance. The key insight is that simply using the right algorithm tools and reasoning path is not sufficient to be correct.

This signals that while conversational AI is ready to add value in specific areas, the vision of a fully autonomous, intent-driven network remains a significant challenge. Further research progress requires a holistic approach that addresses the entire reasoning pipeline, from planning and decomposition to execution and factual grounding.

\bibliographystyle{IEEEtran}
\bibliography{main.bib}

@ARTICLE{10148954,
  author={Zou, Xiaofeng and Li, Kenli and Zhou, Joey Tianyi and Wei, Wei and Chen, Cen},
  journal={IEEE Communications Standards Magazine}, 
  title={{Robust Edge AI for Real-Time Industry 4.0 Applications in 5G Environment}}, 
  year={2023},
  volume={7},
  number={2},
  pages={64-70},
}

@misc{Omdia,
  author       = "Tomasi, Pablo",
  title        = "Private {LTE and 5G} Network Enterprise Survey Insights",
  year         = "2024",
}

@misc{Nvidia,
  title        = "https://www.nvidia.com/en-gb/glossary/ai-ran/",
  year         = "2025",
}

@ARTICLE{RAG6G,
  author={Tang, Yun and Guo, Weisi},
  journal={IEEE Communications Magazine}, 
  title={Automatic Retrieval-Augmented Generation of {6G} Network Specifications for Use Cases}, 
  year={2025},
  volume={63},
  number={4},
  pages={95-102},
}

@ARTICLE{Graph,
  author={Xiong, Yang and Zhang, Ruichen and Liu, Yinqiu and Niyato, Dusit and Xiong, Zehui and Liang, Ying-Chang and Mao, Shiwen},
  journal={IEEE Wireless Communications}, 
  title={When Knowledge Graph Meets Retrieval Augmented Generation for Wireless Networks: A Tutorial and Case Study}, 
  year={2025},
  volume={},
  number={},
  pages={1-9},
}

@ARTICLE{AgentInterface,
  author={Zhang, Ruichen and Du, Hongyang and Liu, Yinqiu and Niyato, Dusit and Kang, Jiawen and Sun, Sumei and Shen, Xuemin and Poor, H. Vincent},
  journal={IEEE Network}, 
  title={Interactive AI With Retrieval-Augmented Generation for Next Generation Networking}, 
  year={2024},
  volume={38},
  number={6},
}

@ARTICLE{RRM,
  author={Zeeshan, Hafiz Muhammad Ali and Umer, Muhammad and Akbar, Muhammad and Kaushik, Aryan and Jamshed, Muhammad Ali and Jung, Haejoon and Hassan, Syed Ali},
  journal={IEEE Communications Magazine}, 
  title={LLM-Based Retrieval-Augmented Generation: A Novel Framework for Resource Optimization in 6G and Beyond Wireless Networks}, 
  year={2025},
  volume={63},
  number={10},
}

@ARTICLE{RFSensingGPT,
  author={Zakir Khan, Muhammad and Ge, Yao and Mollel, Michael and Mccann, Julie and Abbasi, Qammer H. and Imran, Muhammad},
  journal={IEEE Transactions on Cognitive Communications and Networking}, 
  title={RFSensingGPT: A Multi-Modal RAG-Enhanced Framework for Integrated Sensing and Communications Intelligence in 6G Networks}, 
  year={2026},
  volume={12},
  number={},
  pages={298-311},
  doi={10.1109/TCCN.2025.3558069}}

@ARTICLE{FLGPT25,
  author={Yuan, Jinsheng and Tang, Yuan and Guo, Weisi},
  journal={IEEE Vehicular Technology Conference (VTC)}, 
  title={{RAG-based User Profiling for Precision Planning in Mixed-precision Over-the-Air Federated Learning}}, 
  year={2025},
}

@ARTICLE{AutoML25,
  author={Tang, Yuan and Zou, Mengbang and Srinivasan, Udhaya and Umealor, O and Kevogo, D and Scott, Ben and Guo, Weisi},
  journal={IEEE Global Communications Conference (Globecom)}, 
  title={{Building AI Service Repositories for On-Demand Service Orchestration in 6G AI-RAN}}, 
  year={2025},
}

@ARTICLE{AgentBench,
  author={Xiao, Liu and et al.},
  journal={International Conference on Learning Representations (ICLR)}, 
  title={AgentBench: Evaluating LLMs as Agents}, 
  year={2024},
}

@ARTICLE{10758700,
  author={Chatzistefanidis, Ilias and Leone, Andrea and Nikaein, Navid},
  journal={IEEE Networking Letters}, 
  title={Maestro: LLM-Driven Collaborative Automation of Intent-Based 6G Networks}, 
  year={2024},
  volume={6},
  number={4},
  pages={227-231},
  doi={10.1109/LNET.2024.3503292}}

\end{document}